# Achieving a quantum smart workforce

Clarice D. Aiello, Department of Electrical and Computer Engineering, University of California, Los Angeles, Los Angeles, CA, 90095 USA

D. D. Awschalom, Pritzker School of Molecular Engineering, University of Chicago, Chicago, IL 60637, USA; Department of Physics, University of Chicago, Chicago, IL 60637, USA; Center for Molecular Engineering and Materials Science Division, Argonne National Laboratory, Lemont, IL 60439, USA

Hannes Bernien, Pritzker School of Molecular Engineering, University of Chicago, Chicago, IL 60637, USA

Tina Brower-Thomas, Howard University, Research Professor, Howard University Graduate School, Washington, DC 20059, USA; Center for Integrated Quantum Materials, Howard University, PI; Washington, DC 20059, USA

Kenneth R. Brown, Departments of Electrical and Computer Engineering, Chemistry and Physics, Duke University, Durham, NC 27708, USA

Todd A. Brun, Department of Electrical and Computer Engineering, University of Southern California, Los Angeles, California 90089, USA

Justin R. Caram, UCLA Department of Chemistry and Biochemistry and Center for Quantum Science and Engineering, University of California – Los Angeles, Los Angeles, California 90095, USA

Eric Chitambar, Department of Electrical and Computer Engineering, University of Illinois, Urbana-Champaign, IL 61801, USA

Rosa Di Felice, Department of Physics and Astronomy, University of Southern California, Los Angeles, CA 90089, USA

Michael F. J. Fox, JILA, National Institute of Standards and Technology and University of Colorado, Boulder, Colorado 80309, USA; Department of Physics, University of Colorado, Boulder, Colorado 80309, USA

Stephan Haas, Department of Physics and Astronomy, University of Southern California, Los Angeles, California 90089, USA

Alexander W. Holleitner, Technical University of Munich and Munich Center for Quantum Science and Technology (MCQST), Garching 85748, Germany

Eric R. Hudson, UCLA Center for Quantum Science and Engineering, University of California – Los Angeles, Los Angeles, California 90095, USA

Jeffrey H. Hunt, The Boeing Company, El Segundo CA 90245, USA

Robert Joynt, Department of Physics, University of Wisconsin-Madison, Madison, WI 53706, USA

Scott Koziol, Department of Electrical and Computer Engineering, Baylor University, Waco, TX 76798, USA

H. J. Lewandowski, Department of Physics, University of Colorado, Boulder, Colorado 80309 USA and JILA, National Institute of Standards and Technology and University of Colorado, Boulder, Colorado 80309, USA

Douglas T. McClure, IBM T.J. Watson Research Center, Yorktown Heights, NY 10598, USA

Jens Palsberg, Department of Computer Science, University of California – Los Angeles, Los Angeles, California 90095, USA

Gina Passante, Department of Physics, California State University, Fullerton, Fullerton, CA, 92831, USA

Kristen L. Pudenz, Lockheed Martin Aeronautics, Fort Worth, Texas, USA

Christopher J.K. Richardson, Laboratory for Physical Sciences, University of Maryland, College Park MD 20740, USA

Jessica L. Rosenberg, Department of Physics and Astronomy, George Mason University, Fairfax, VA 22030, USA

R. S. Ross, HRL Laboratories, LLC, Malibu, California 90265, USA

Mark Saffman, Department of Physics, University of Wisconsin-Madison, Madison, WI, 53706, USA

M. Singh, Department of Physics, Colorado School of Mines, Golden, CO 80401, USA

David W. Steuerman*, The Kavli Foundation, Los Angeles, CA 90230, USA

Chad Stark, The OSA Foundation and The Optical Society, 2010 Massachusetts Avenue, NW, Washington, DC 20036, USA

Jos Thijssen, Kavli Institute of Nanoscience, Delft University of Technology, Building 22 Lorentzweg 1 2628 CJ Delft, The Netherlands

A. Nick Vamivakas, The Institute of Optics, University of Rochester, Rochester, NY 14627 USA

James D. Whitfield, Department of Physics and Astronomy, Dartmouth College, Hanover, NH 03755, USA

Benjamin M. Zwickl, School of Physics and Astronomy, Rochester Institute of Technology, Rochester, NY 14623, USA

## ABSTRACT

Interest in building dedicated Quantum Information Science and Engineering (QISE) education programs has greatly expanded in recent years. These programs are inherently convergent, complex, often resource intensive and likely require collaboration with a broad variety of stakeholders. In order to address this combination of challenges, we have captured ideas from many members in the community. This manuscript not only addresses policy makers and funding agencies (both public and private and from the regional to the international level) but also contains needs identified by industry leaders and discusses the difficulties inherent in creating an inclusive QISE curriculum. We report on the status of eighteen post-secondary education programs in QISE and provide guidance for building new programs. Lastly, we encourage the development of a comprehensive strategic plan for quantum education and workforce development as a means to make the most of the ongoing substantial investments being made in QISE.

## INTRODUCTION

The meteoric rise of interest in Quantum Information Science and Engineering (QISE) is built on the more than a century old theory of quantum mechanics, decades of fundamental research breakthroughs in quantum physics broadly defined, and recent revolutionary demonstrations of quantum simulators (1), sensors (2), networks (3), and computers (4). Thanks to the ingenuity and perseverance of many, we can now safely say that Richard Feynman's oft-recited quote, "I think I can safely say that nobody understands quantum mechanics" is long since obsolete. There is already a substantial community of QISE researchers, and an ever-growing number of scientists and engineers who are moving into the field. In fact, applications of quantum mechanics comprise some of the most promising areas of the physical sciences and engineering today.

The many successes of QISE and its seemingly endless potential for basic science and technology has led to a large syndicate of stakeholders. Currently, QISE is strongly supported by national funding agencies, a myriad of industry players, private philanthropy, and scientific societies. Many students at a variety of education levels, as well as active members of the global workforce, are interested in learning QISE to be part of this potentially radical transformation of science and technology. Consequently, many researchers, policy makers, and universities have created or are exploring dedicated programs to educate the quantum scientists and engineers of tomorrow.

Building new education programs at the intersection of many disciplines, like QISE, is a complex process. Fortunately, the global quantum community is well positioned to face such challenges. Because of the establishment of the United States National Quantum Initiative (5,6) the European Union's Quantum Flagship (7), large efforts in China (8), and Japan (9), investments from industry leaders, and their growing appetite for a talented "Quantum Smart" workforce, as well as an actively engaged academic community, QISE has a solid foundation to meet these demands.

Towards this goal, in November of 2019, a symposium gathered approximately 50 US and European QISE experts from both industry and academia to discuss post-secondary education with an emphasis on Master's programs in QISE (the authors believe many of the lessons herein would apply at other educational levels). Figure 1 depicts many of the participants of the two-day workshop focused on panel discussions, open debate, and candid conversations. At the conclusion of the event, it was decided that a synthesis of the pre-meeting data gathered, discussion and opinions shared during the symposium, and targeted post-meeting work from select attendees would be of value to the QISE community at large. D.W. Steuerman was nominated as the editing author and the coordinator of the group in charge of presenting the attendees' perspectives. Efforts to this effect started immediately following, and this document represents the views of its contributors.

In this article, we identify many of the challenges of educating the quantum workforce at the post-secondary level and what is and can be done to overcome them. The first half of the article is focused on defining QISE and the needs of the emerging quantum industry as well as the challenges of formulating a relevant, inclusive QISE curriculum. In the second half of the article, we provide detailed information on the status of 18 QISE programs (a subset sampled from the US and Europe) with emphasis placed on addressing the issues identified in the first half. We conclude by identifying lessons learned from other convergent education programs, factors helpful in establishing a program, and finally call for the development of strategic plan for quantum education and workforce development. Our primary objective is that by sharing this collective perspective from a wide variety of leading scientists, engineers, and educators, we can more deliberately and thoughtfully shape the future of quantum education.

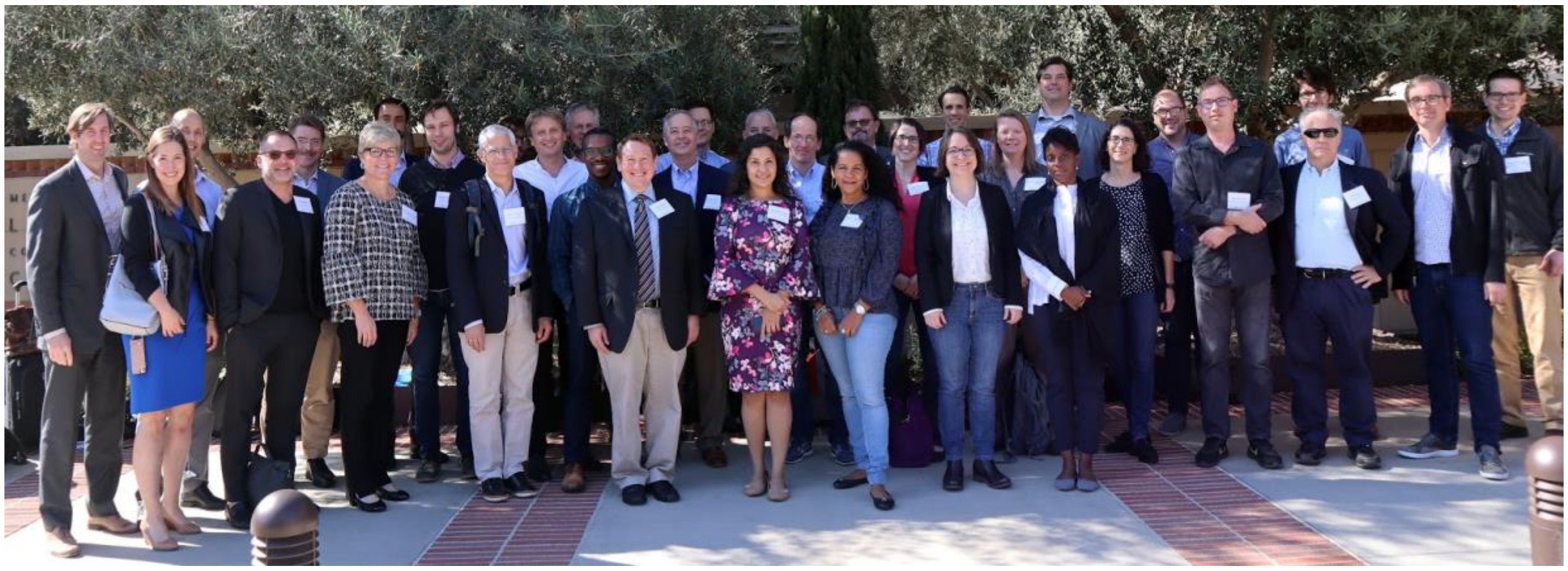

**Fig. 1 Attendees of the Kavli Futures Symposium: Achieving a Quantum Smart Workforce.** This event took place on November 4th and 5th of 2019 at the University of California, Los Angeles campus.

## DEFINING QUANTUM INFORMATION SCIENCE AND ENGINEERING

Quantum Information Science and Engineering seeks to utilize quantum phenomena to create enhanced capabilities for information processing, communication, and sensing. Although the theory of quantum mechanics underpins all physical phenomena, our everyday experience is anchored in a classical world where explicitly quantum phenomena are obscured. The separation between quantum and classical phenomena is evident, for example, in classical computers that store data as binary bits. The bits are classical and take on the values of 0 or 1, while the underlying objects that are used to build classical bits, be they electrons or magnetic spins, are fundamental quantum objects that can in principle be in a superposition of more than one state at a time. In a quantum computer, phenomena such as superposition and entanglement are manifest, leading to qubits that can be in superpositions of 0 and 1, and can be entangled with each other, thereby providing a much richer set of possibilities for storing, processing, and securing information than in a classical computer (10). QISE seeks to develop enhanced capabilities using technologies that exhibit and leverage or control manifestly quantum phenomena.

## INDUSTRY PERSPECTIVES ON THE WORKFORCE

In order to take full advantage of the quantum opportunity and realize a new age of scientific discovery and technology, it is imperative that emerging QISE post-secondary education programs seek input from industry partners (11). While initially pioneered by academic research groups and a handful of corporate research labs (12-15), optimism that early practical applications of QISE may be on the horizon has recently led to rapid growth in the number and diversity of companies participating in the field. Today, companies big and small are already heavily engaged in quantum technology development, and these early-mover organizations are among the most highly desired destinations for future graduating students of QISE programs. While some companies focus on developing hardware and/or software pertaining to core QISE technologies (computing, sensing, or communications), others focus on exploring potential applications in finance, chemistry, medicine, materials development, or other specialized areas.

It is clear that the breadth of the above interests within the quantum industry cannot be met by a single educational prescription nor a single level of talent. There will be a need for a wide variety of expertise and education levels to create a balanced technical workforce like that seen in other professional scientific and engineering fields. Furthermore, the rapid pace of innovation in the field of QISE poses challenges for the development of standardized classroom- and lab-based curricula. Specific types of instruments, quantum systems, and even concepts that are *en vogue* today may become obsolete tomorrow as the field evolves. Students getting started in QISE should expect to be continually learning throughout their career, potentially even more than they might need to in other science, technology, engineering, and mathematics (STEM) fields. These considerations suggest that the most important component of an education in QISE is the development of a strong set of fundamental concepts and skills that will provide a foundation for future learning. Practical training in the form of internships, capstone projects, and/or laboratory-based courses is consequently vital: it serves to reinforce and refine understanding of abstract concepts presented in the classroom; offers students a taste of the type of work they might do if they pursue a career in QISE; and allows them to develop and demonstrate skills sought by prospective employers.

Some of the skills and knowledge areas that were highlighted by industry leaders as important ingredients for contributing to industry quantum projects are the following (in no particular order): linear algebra, programming and software engineering fundamentals, various lab-related skills (electronics, vacuum and cryogenic techniques), statistics and statistical mechanics, and an introductory college-level understanding of quantum information and quantum physics, which is especially suitable for students in computing and engineering fields who enter QISE companies. Some industry members pointed out that as quantum hardware advances and becomes more modular, more opportunities are arising for specialists in nontraditional QISE fields to have an impact as well; professional education or certificate opportunities could have positive effects here. Lastly, it must be emphasized that all industry leaders spoke about the importance of being able to succeed on team projects and effectively communicate technical ideas to a broad audience.

Although QISE companies are all in some way leveraging quantum phenomena for commercial applications, defining the QISE workforce is subtler. Quantum education is not just relevant for PhD programs at elite universities, but needs to be considered from the earliest years of science and engineering education (16). Within a company's workforce, some roles require very advanced quantum knowledge and skills, often at the PhD level, while other roles include "non-quantum" engineers, software developers, and technicians who contribute immensely to the designing, making, selling, and supporting of products. These other "non-quantum" roles are essential, yet require much less formal training in quantum. Additionally, there will be an ever-increasing need for effective and scientifically accurate communication by those who describe new quantum technologies to the public. When considering the response of higher education to train this workforce, the full range of job types must be considered. This article focuses on training technical experts, but programs to improve quantum literacy need to be thoroughly considered as well.

## BROADENING PARTICIPATION AND INCLUSION

Broadening participation is a challenge that has not been overcome in the STEM disciplines (17). If the QISE community is going to meet the workforce needs in this area, it needs to utilize all of our human capital. Moreover, teams that are diverse are more productive (18). If QISE expands rapidly without addressing the barriers to access and inclusion that currently exist, these problems will be solidified into the next generation. The community must consider the mentoring, culture, and support within our interdisciplinary programs to remove barriers and help students navigate complex cultural differences. For students not deeply embedded in the cultures of QISE-related departments, this interdisciplinary landscape may prove challenging. The QISE community must create programs that value differing expertise and perspectives from the beginning to increase accessibility.

To date, the strongest involvement in QISE has come from disciplines that have some of the lowest representation of women within STEM. According to the 2017 NSF Science and Engineering Indicators, women earned a smaller percentage of Bachelor's degrees than men in the primary quantum-related disciplines: computer sciences (19%), engineering (22%), mathematics and statistics (42%), and physical sciences (40%). In the US, women account for 60% of undergraduate degrees, 22% of engineering degrees, and yet only 13% of the engineering workforce (19). As new courses and programs are developed, a crucial first step is to collect data on how these courses are developing students' self-efficacy (belief they can succeed), providing recognition, growing their interest, and developing their identity as a member of the QISE community. These factors are very influential in career decision-making within STEM, and specifically the physical sciences (20,21). Moreover, QISE courses should be designed to highlight the contributions of underrepresented scientists to the field. While the QISE community is still nascent, emphasizing diversity up front, rather than as an afterthought, is an essential step forward. Some organizations and events are already forming, such as the Women in Quantum Development Symposium (22).

Similar considerations are necessary to promote inclusion of diverse racial and ethnic groups. The 2017 NSF Science and Engineering Indicators show that students who identify as Hispanic, Latinx, Black or African American account for a much higher percentage of awarded degrees at the Associate's degree level than at the Bachelor's degree level: computer sciences (27%), engineering (27%), math and statistics (38%), and the physical sciences (35%). The community needs introductory-level courses or modules with QISE content that could be taken during the first two years of an undergraduate degree, rather than deferring such material to more advanced courses that are accessible to fewer students. Such courses could serve as a gateway to a more advanced degree and raise awareness about the opportunities and potential career paths. At the PhD level, programs such as the American Physical Society Bridge Program (23) and the Inclusive Graduate Education Network (24), are shaping how PhD programs approach admissions, retention, and professional development with the aim of increasing participation of underrepresented racial and ethnic minority groups. A culture of inclusion, mutual respect, and appreciation of culture differences should be an integral part of all education. The QISE community should build on these successes by taking advantage or creating new quantum focused programs.

## COMMENTS ON CURRICULA, CLASSES, AND LABS

As QISE education programs take shape and evolve, they will be faced with a variety of pedagogical challenges to meet wide ranging industry needs given the breadth of commercial and research activities and the pace at which they are maturing. Programs will be expected to address complex subject matter and provide opportunities for hands-on engagement with sophisticated, ever-changing instrumentation. Adding to the challenge, these demands must be met with a broad diversity of subject matter expertise of enrollee backgrounds while providing an environment conducive to team projects. Below are two principles that emerged during our discussions to aid achieving these lofty goals with Master's programs.

**Creating one core course, or a short series of courses, dedicated to QISE principles and applications is crucial when providing students with diverse educational backgrounds a common language.** There was widespread agreement that this will not be a traditional quantum mechanics course routinely offered in a physics department. It must be designed for non-physicists and have pre-requisites commensurate with students with undergraduate degrees in any technical field such as physics, engineering, computer science, chemistry, and math. It will likely be heavily based on a linear algebra approach to quantum mechanics. As companies move toward manufacturing of quantum technologies, the need will increase for engineers, software developers, and technicians who are familiar with the core ideas, but who are primarily valued for their traditional expertise in engineering, computing, or manufacturing. Creating one or more introductory-level QISE courses would directly meet this need, as well as serve as a foundation for more advanced study of QISE topics.

**The value of experimental skills related to quantum technologies can be equally, or even more, important for entering the workforce than courses in complex quantum theory at the Master's level.** However, this hands-on learning is often omitted from programs as it is resource intensive, which is a barrier to scaling class size. The necessary resources include equipment, infrastructure, and student and instructor time. To ensure these resources are being used efficiently, program designers should consider the specific skills and learning outcomes they are aiming to

achieve with every laboratory course, research experience, internship, co-op, and project-based class and how the goals of each course build to meet overall program goals. There were several broad categories of desired skills identified by the participants including (1) Technical knowledge (2) Data analysis (3) Experimental and engineering design (4) Modeling of experiments (5) Troubleshooting (6) Documenting and reporting (7) Programming and coding, including software for experimental control (8) Interdisciplinary teamwork.

A potential advantage of such an experiential-based approach is that the relevance of these skills is not specific to the QISE context, but is transferable to a variety of industries beyond quantum. This both provides educational security in the event that quantum industry does not grow as expected and better prepares students for work in a successful and rapidly changing quantum industry. Industry experts were explicit that courses should be broad and not focus solely on any one QISE architecture or technique. The field is still young, and it is unknown which architectures and techniques will ultimately be the most broadly used.

## STATUS OF SAMPLED UNIVERSITY PROGRAMS

In preparation for the symposium, attendees from eighteen universities completed surveys describing their quantum education programs (fifteen in the United States and three in Europe). Given early feedback, the focus was on Master's programs. Contributions included both public and private universities, and span from large to small institutions (nearly 50k students to slightly more than 5k). Based on the survey results, we provide a snapshot of the programs in Table 1 and identify emerging themes below. The programs are in different states of readiness and all build off of pre-existing courses: four are already operating (indicated by an * in Table 1), eight plan to start in 2020, one aims to start in 2021, and the remainder are in the planning stage.

| Institution | # of Courses | # of Students per year | Program Length (years) | Lab Experience / Internship / Project |
|---|---|---|---|---|
| **Master's Program** | | | | |
| Colorado School of Mines | 10 | 30 | 1 to 2 | ● ○ ● |
| Delft University of Technology* | 9 | 75 | 2 | ● ● ● |
| Duke University* | 10 | 5 | 2 | ○ ● ○ |
| ETH Zurich, Switzerland* | 11 | 25 | 2 | ● ● ● |
| George Mason University | 11 | 20 | 2 | ● ● ○ |
| Purdue University | 8 | 15 to 20 | 1.5 | ● ○ ● |
| Rochester Institute of Technology | 4 to 6 | 20 | 1 | ● ○ ● |
| TU Munich and LMU | 11 | 50 | 2 | ● ○ ● |
| University of Arizona | 8 to 10 | 25 | 1 to 2 | ● ○ ● |
| UCLA | 10 | 16 to 32 | 1 | ● ● ● |
| UIUC | 6+ | 10 | TBD | ● ○ ○ |
| University of Rochester | 7 | 15 to 25 | TBD | ● ○ ○ |
| University of Southern California | 8 | 10 to 15 | 1.5 to 2 | ● ○ ○ |
| University of Wisconsin-Madison* | 9 to 10 | 20 to 25 | 1 | ● ○ ● |
| **Certificate Program** | | | | |
| Dartmouth College | 6 | 5 to 10 | 1 to 2 | ○ ○ ○ |
| MIT* (online) | 10 | 250 | 1 | ○ ○ ○ |
| University of Chicago (retraining) | 1 | 30 | 1 *(week)* | ● ○ ○ |
| **PhD Program** | | | | |
| Harvard University | 10 | 10 | 5 | ● ○ ● |

**Table 1 Characteristics of sampled Quantum Science & Engineering programs**. Column 1 is the Institution sorted by degree level, column 2 represents the number of courses required to complete a degree, column 3 is the estimated size of the program cohort, column 4 is the expected time to degree, and column 5 indicates other features of the respective programs. *Lab Experience* implies hands-on experience or physical demonstrations, *Internship* means extended time working with an industrial partner, and *Project* means there is a dedicated course for research or a thesis. An asterisk indicates that the program was running as of Fall 2019.

These programs all have the ambition to provide quantum education at the forefront of knowledge. The most common goals are to educate current students towards research or a job in industry; to educate recent graduates and people already working in industry; and to provide a truly interdisciplinary education. Some programs cover only theoretical material, while most programs cover a mix of theory and laboratory topics. Many of those labs include hands-on experience with quantum equipment in areas such as quantum optics, quantum sensing, and quantum materials. Most of the programs cover quantum computing, many with exercises that students will run on quantum computers that are either local or web accessible. By design of the workshop most attendees discussed programs geared toward the Master's degree, while others described PhD degrees, enhancement of existing undergraduate degrees, certificates for continuing education of industry professionals, and Summer schools. While most of the degrees give traditional on-campus experiences, some are online, and some are a mix. Programs unanimously reported that tenured and/or tenure-track faculty will teach almost all the courses, often as part of a regular teaching load. In most cases, courses within quantum programs will not be exclusive to program enrollees, but open broadly to other majors. The number of students in each cohort varies from 10 students in a program to 250 students in an online program. Once all the surveyed programs are operating, they could educate a total of ~500 students per year and many are interested in increasing their number of students, provided there is sufficient demand and that various resource constraints can be addressed.

There were many shared concerns reported by program representatives prior to the workshop. Chief among them was whether there is sufficient demand from students for the degree and sufficient demand from industry for graduated students with a terminal Master's degree.

Additionally, many programs reported uncertainty about the industry needs, which, if known, could help shape the programs (11). Similarly, there was significant concern about the difficulties in accepting students from a broad variety of majors into a program.

Unsurprisingly for interdisciplinary programs, there is variation in what department(s) or units the programs reside in on campus. Figure 2 depicts the distribution of sampled programs across departments. Almost half of the programs are housed in a physics department, while 18% are housed in an engineering department. Additionally, some of the remaining programs are offered jointly by multiple entities or reside in an interdisciplinary entity. Many programs reported concerns about how to navigate university administration in order to establish a program, how to ensure that a sufficient number of faculty can teach the courses, and how to manage a complex interdepartmental (and sometimes inter-college) collaboration with regard to course offerings and infrastructure. Below, we highlight three university programs who have sufficiently overcome these concerns and have launched or will do so in 2020.

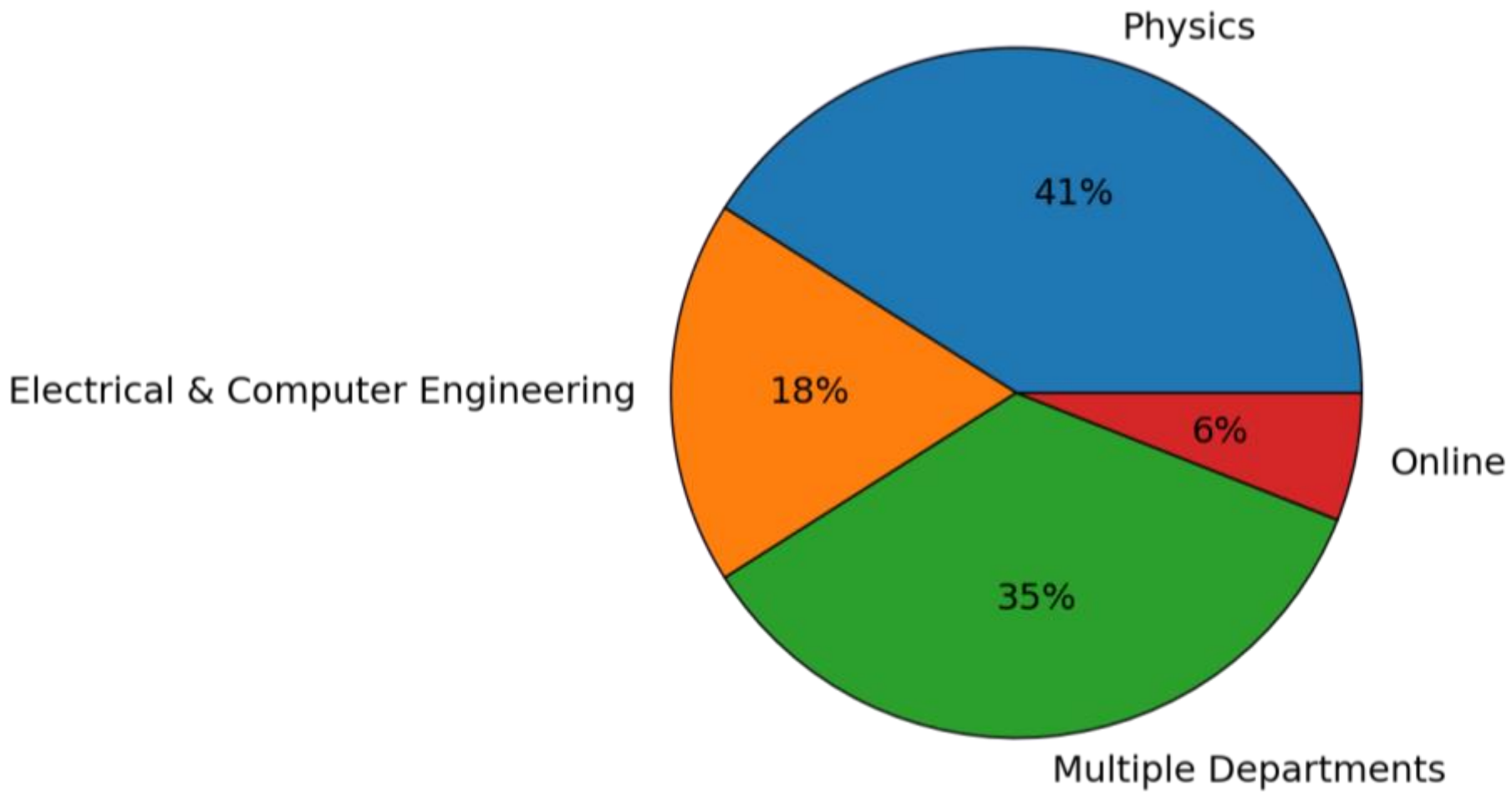

**Fig. 2 Distribution of Departmental Affiliation of Sampled Quantum Science and Engineering Programs.** Many proposed programs reside in Physics Departments solely and an almost equal number are across multiple departments. The sole online offering resides within the office of the Vice President of Online Education.

### Case Study (UW-Madison)

Quantum information has been taught at the undergraduate and graduate levels at University of Wisconsin-Madison (UWM) since 2008. During this time, courses have been offered in the Physics and Computer Sciences departments. In the Fall of 2019, the Physics department began offering a professional Master’s degree, M.S. Degree in Physics-Quantum Computing (MSPQC). This is a one-year program with two semesters of classroom instruction, followed by a summer laboratory course and an independent study project. The classroom instruction component includes existing physics courses and a two-course sequence in quantum information specially developed for the MSPQC degree.

The MSPQC specialized two course sequence is comprised of “Introduction to quantum computing” and “Advanced quantum computing.” Both courses are three hours per week of classroom instruction during a 15-week semester. The first provides the necessary quantum mechanics background needed to understand qubits, quantum gates, and quantum circuits. The second course introduces the quantum theory of open systems using density operators and treats decoherence and quantum processes. A survey of leading experimental implementations is given including trapped ions, neutral atoms, photons, semiconductor quantum dots, and superconducting circuits, as well some coverage of quantum sensors. This sequence is followed by a summer laboratory course that includes experimental demonstration of basic quantum phenomena with photons, atoms, ions, and superconductors.

### Case Study (Colorado School of Mines)

Colorado School of Mines (Mines) has offered an “Introduction to Quantum Computing” course in the Physics department at the undergraduate and graduate levels since 2017. In the spring of 2020, an M.S. degree (thesis and non-thesis options) and a 12-credit certification program in Quantum Engineering was approved by the administration. Both the degree and the certificate will be offered starting Fall 2020. The program has been developed collaboratively by five academic departments at Mines (Physics, Electrical Engineering, Computer Science, Applied Mathematics and Statistics, and Metals and Metallurgical Engineering) and is aimed at attracting students from disparate majors.

In addition to eight electives, chosen from existing courses offered at Mines that are relevant to quantum engineering, four new ‘core’ courses have been developed for this degree. The first of these – Fundamentals of Quantum Information – is aimed at serving as a broad introduction to quantum information science to students from different educational backgrounds. In this course, the basic structure of quantum mechanics (Hilbert spaces, operators, wave functions, entanglement, superposition, time evolution) will be presented, as well as a number of important topics relevant to current quantum hardware (including oscillating fields, quantum noise, and more). The second course, Quantum Many Body Physics, is focused on entanglement as a central theme. Topological ordered quantum matter and its relevance to quantum computing will also be introduced. In the third course, focused on Quantum Programming, students will receive an in-depth education in quantum algorithms and their design, and then break into teams to learn the application programming interface of a commercially available quantum computing system. The fourth is a laboratory course, focused on providing hands-on experience with technologies used in quantum information applications. In addition to a survey of leading quantum computing platforms, this course will include hands-on experience with high-frequency measurement systems, the latest techniques for accuracy-enhanced automated microwave measurements, low-temperature measurement techniques, low noise measurements, and common devices used in quantum computing and sensing. For a thesis-based M.S. degree, a two to three semester long research project with a local research group or an industrial partner will be required along with the necessary coursework. The certification program requires 12 credits of courses from the core courses and electives described above. The pre-requisites of the courses are such that students pursuing their undergraduate degrees from multiple majors will be eligible for this program. The M.S. programs (both thesis and non-thesis track) are one-year programs requiring 30 credits. For the M.S.-thesis based program, 9 of these 30 credits are dedicated to research.

### Case Study (Technical University Munich and Ludwig-Maximilians-University)

In Munich, a Master's degree on "Quantum Science & Technology" (QST) starts in the Fall of 2020 based on joint courses at the Technical University Munich (TUM) and the Ludwig-Maximilians-University (LMU). The program is centered at the physics departments with lectures from the electrical engineering, mathematics, computer sciences, and chemistry departments. The two-year program starts with a one-year (two semester) classroom instruction with lectures and advanced laboratory training at the participating institutions, followed by a one-year study project (master project). During the classroom period, there are two mandatory courses, one on "QST Experiment: Quantum Hardware" and one on "QST Theory: Quantum Information". Both courses are four hours with an additional two hours of exercises per week, and they cover the fundamental aspects from quantum science and technology, including superconducting and semiconducting quantum circuits, atom and ion quantum gases, quantum sensing, communication, and simulation, as well as theoretical concepts of entanglement, non-locality, dense coding, quantum teleportation, quantum cryptography, and quantum information theory. Following up these fundamental courses, professors from TUM and LMU offer a wide-spread and balanced set of special lectures on advanced topics of QST. In the second year, the research-oriented Master's projects take place. The students work on scientific projects with the experimental and theoretical research groups of both TUM and LMU. The program aims to educate researchers who continue as PhD-students in quantum science and technology, but also as professionals working in the high-tech industry.

## LESSONS LEARNED FROM OTHER PROGRAMS

Starting new educational programs in the interdisciplinary field of QISE is a daunting task. To this end, it is insightful to examine other interdisciplinary programs that have already launched and attempt to learn from their experiences. Some analogous programs include biomedical engineering, materials science, data science, robotics, and bioinformatics. The following are insights gained from building new programs in bioinformatics.

Bioinformatics, as a now well-established discipline that combines biology, computer science, mathematics and medicine, a decade ago faced similar challenges that new QISE education programs will encounter. As bioinformatics matured, two main approaches were taken to develop new programs. The first one was an extension of existing computer science training that would include several electives in biology and medicine. This "add-on approach" can build on the strengths and rigor of an existing program while adding new opportunities for the students. The second approach taken was a completely new program that would combine all different disciplines in one interdisciplinary curriculum. This "all-new approach" has the attraction of creating a new discipline that is not bound to the conventions of existing ones, but at the same time comes with many challenges. One has to carefully evaluate if the all-new approach truly adds value when compared to the add-on approach, given the likely increased administrative burden of building a new program that will cross multiple deans and schools and the difficulties in training students with such diverse educational backgrounds.

Similar considerations must be made for new QISE programs. Many new programs are housed in physics departments (see Figure 2) and extend their curricula by adding courses in engineering, computer science, and material science while others are positioned as new disciplines in between these departments. Both approaches can succeed, and which one is better suited depends on the local circumstances. Crucial for success is strong support and buy-in from all involved departments.

## STEPS TOWARDS BUILDING A PROGRAM

Present at the workshop were many quantum-focused faculty, university administrators, industry scientists who support academic research, and funders. The group also included expertise in building interdisciplinary programs outside of QISE, those who have already launched quantum programs, and those about to embark on such a journey. The consensus that was reached after many discussions was that a successful program would greatly benefit from having taken many of the following actions and possessing a clear path to address those that are missing:

**Gathering subject-matter experts:** This is to ensure a sufficient number of qualified personnel for teaching necessary courses and to provide an adequate number of diverse projects if research experience is included in a program. Following several principles for forming this group of experts could stave off problems later. First, pedagogy experts should be included in this initial cohort, so that educational best practices and assessment methods are incorporated in the program from the beginning. Second, assigning multiple faculty to develop each course could help the program stay true to its vision. This will also ensure continuity in subject matter and relationship to other courses when course instructors change. Third, it is important to ensure that faculty from diverse departments are included in this group. This ensures that considerations unique to every discipline are accounted for in designing the curriculum and will prove to be important in attracting and retaining students from different departments.

**Establishing mechanisms to ensure institutional commitment:** A cross-disciplinary, program-level innovation requires consistent institutional support. This support is necessary for informing hiring decisions, ensuring teaching and service assignments for the program do not conflict with departmental duties, equipment support (if relevant), and resource allocation for program assessment and improvement including career-related assistance. Consulting with colleagues who have experience creating successful cross-disciplinary programs in the specific school(s), regular meetings with everyone potentially affected by the program (an informal quantum coffee hour could suffice), and incorporating the administration's vision for the school in the program are some important strategies for achieving this.

**Developing strong relationships with industry:** One of the main motivations for training a 'quantum workforce' is to meet industry needs. To adequately assess and address industry needs, a close relationship with industry, ideally involving student and employee exchange, is needed. Even something relatively simple such as providing access to industry technology, such as cloud quantum computing time not available to the public, can bolster the strength and industrial preparedness of QISE graduates. Particularly important is locating effective points of contact in large industrial organizations in which QISE may be only a small part. Often, large companies are willing to send a representative to an educational institution to introduce students to the employment opportunities that the company can provide. Incorporating industrial internship experiences, introducing thesis topics chosen with industrial input, assigning committee members from industry, and accepting part-time students who work in the relevant industry are all excellent methods of building a program that is relevant to industry. This aspect has the potential to address several possibly problematic areas of a quantum engineering degree (especially an M.S. degree).

First, with industry support, it can reduce costs associated with an extra year or two of college. Second, it can prevent the degree from becoming a token one, instead, providing access to jobs that a B.S. would not. Finally, particularly for universities without strong experimental research programs in QISE, it will prevent the course material, especially if laboratory instruction is involved, from becoming obsolete by constantly allowing students exposure to cutting edge equipment used in industry. Identifying contacts at companies can be facilitated by engaging with the Quantum Economic Development Consortium (QED-C), which includes a broad cross-section of the quantum industry and whose mission includes quantum workforce development.

**Acquiring resources for laboratory courses:** Teaching some courses, especially courses with a laboratory component, can be especially expensive in this field. Resources can be acquired via several means. Federal agencies offer some opportunities for funding for innovations in education. Some private organizations, like the Reichert Foundation and the Research Corporation for Science Advancement, do the same. In addition, most institutions have some means of funding laboratory upgrades (this is also an area where institutional commitment is important). Finally, relationships with industry can potentially be leveraged for cost sharing of equipment.

**Setting up assessment and dissemination mechanisms:** Timely assessment ensures that the program receives feedback and improves. Dissemination is critical so that QISE community members all learn together and do not make the same mistakes. Having experts in research-based curriculum development included during the generative stages of the program will ensure that meaningful assessment is incorporated into a program. All the faculty involved in course creation and delivery must be committed to administering the assessments developed by discipline-based education researchers. The findings of these assessments can be disseminated through education related journals and conferences or publicly accessible repositories maintained by various institutions.

**Hiring a Faculty Cluster:** QISE research and teaching has been historically initiated primarily in physics departments. The growth of QISE activity is being mirrored in the expansion of faculty appointments in other departments besides physics. One challenge for departments that have not previously engaged in QISE topics is the difficulty of growing new faculty lines in the absence of existing expertise in the department. In this situation, cluster hires that make appointments in several departments simultaneously can facilitate campus-wide growth of the QISE faculty. Designing a cluster hiring process that combines departments with QISE expertise with those where QISE activity is sought but not yet in place can be beneficial, provided that mechanisms are put in place to ensure coordination and consultation between departments throughout the hiring process.

Not all of the above actions are necessary nor sufficient to guarantee success, but such steps will put a program on an excellent trajectory. These kinds of efforts will signal to the faculty, students, as well as industry partners, that there is substantial commitment to the program. This is not a prescription and every university will develop their program differently, but it is crucial that such new and ambitious programs have strong foundations from which to evolve or they risk being fragile in the competitive and rapidly moving space of QISE.

## CONCLUSION AND OUTLOOK

The rapidly increasing interest in quantum technologies from industry, academia, and others requires a parallel rise in QISE education. Creating effective programs at the post-secondary level is a challenging task as there are a myriad of stakeholders, including future employers, educational institutions, and students, each with their own needs and challenges. Future employers, such as those from industry, are requesting QISE-ready graduates even as the field is rapidly evolving. Universities that teach these students face multifaceted difficulties, including limits in: academic talent to teach, access to costly complex physical resources, partnerships with industry, support from university administration, and appropriately architected programs that can support broadly trained and diverse students to produce workforce-ready talent that is superior to traditional program graduates. Many universities are planning to or have already decided to accept the challenge of preparing the QISE workforce.

QISE currently possesses the rare combination of intense interest and investment from a broad segment of industrial and government leaders, substantial support from funders of basic research, and the passion, curiosity, and ingenuity of many in academia. Now is an opportune time to develop an ambitious strategic plan for quantum education and workforce development that appreciates the interconnectedness of all stakeholders, starts QISE education earlier in the academic career, addresses QISE at all education levels for students across academic, gender, racial, and ethnic backgrounds, and embraces the convergent nature of the subject matter. This plan must include an investment in a wide range of institutions, centers, programs and leave open opportunities for international collaboration. On one end of the spectrum, there is a need for graduate programs that are advancing technology and research at the highest-level and provide research experiences and coursework to train future leaders in QISE research and development. However, as quantum technologies become mainstream some level of quantum knowledge will need to be accessible to engineers, software designers, field technicians, and others who are involved in manufacturing and using these quantum technologies. There is a need for broad student access, which could include new introductory-level science and engineering classes that introduce students to the foundations of quantum information science and technology as early as possible in their academic careers (25, 26). Investments in these courses should not be restricted to universities which also have large research centers, but should include community colleges and primarily undergraduate institutions. Emphasizing quantum within these institutions will drastically improve the quantum literacy of all STEM majors

There are already excellent examples of projects and programs that are addressing some of these needs and bolstering the QISE community. There are projects such as the DOE's Advanced Quantum Testbed (27) and Europe's OpenSuperQ (28) and AQTION (29), that will give students and researchers access to state-of-the-art quantum computing systems. The Chicago Quantum Exchange (30) is a powerful example of a regional hub dedicated to quantum innovation and education creating connections between industry, university, and national lab stakeholders. QED-C, which is supported by the U.S. government and industry, aims to grow the quantum industry and its workforce. With continued coordinated investment in science and education, QISE has a genuine opportunity to be an exemplar convergent scientific area that not only moves the frontiers of science and engineering, but also demonstrates what a vibrant 21st century community can look like and accomplish given the right support.

## ACKNOWLEDGEMENTS & NOTES

We gratefully acknowledge discussions with Corey A. Stambaugh (OSTP/NQCO), UCLA's Center for Excellence in Engineering and Diversity (CEED, Catherine Douglas); UCLA Women in Engineering (Audrey Pool O'Neal, Breann Branch, Ashley Fletcher); and Scott Brandenberg and attendee suggestions from Ignacio Cirac (Max-Planck-Institute of Quantum Optics) and Lieven Vandersypen (TU Delft). The authors also wish to recognize the National Science Foundation's efforts to improve the way quantum physics is taught in K-12 by identifying critical concepts for early learners, their focus on academic and industrial connections through the TRIPLETS program and their partnership with the Department of Energy to support Quantum Science Summer Schools (QS3). The authors acknowledge the efforts of The Optical Society (OSA), which through the National Photonics Initiative, worked closely with the academic and industrial QIST community and advocated for the National Quantum Initiative; the Institute of Electrical and Electronics Engineers (IEEE) for their efforts to engage the QIST community broadly (https://quantum.ieee.org/).

## FOOTNOTES

To whom correspondence should be addressed: Email: dsteuerman@kavlifoundation.org

Author contributions: All authors contributed to the discussions and/or provided data that the paper is based upon. C.D.A, D.D.A, H.B, T.B., J.R.C., S.H. A.W.H, E.R.H., R.J, S.K. H.J.L., D.T.M., J.P., K.L.P., J.L.R., M.Sa., M.Si, D.W.S., J.D.W., B.M.Z. wrote the paper.

The authors declare no conflict of interest.